\documentclass[]{iopart}
\usepackage{iopams}
\usepackage{hyperref}
\usepackage{amssymb}
\usepackage{color}
\usepackage{epsfig}
\usepackage{latexsym}
\usepackage{wasysym}

\newcommand{\gw}{gravitational wave }
\newcommand{\gws}{gravitational waves }

\begin{document}

\title[Prospects for joint radio-GW searches]{Prospects for joint radio
telescope and gravitational wave searches for astrophysical transients}
\author{V.  Predoi$^{1}$, J. Clark$^{1}$, T. Creighton$^{2}$, E. Daw$^{3}$, S. Fairhurst$^{1}$, I. S.
Heng$^{4}$, J.  Kanner$^{5}$,  T. Regimbau$^{6}$, P. Shawhan$^{5}$, X. Siemens$^{7}$, P. Sutton$^{1}$, A.
Vecchio$^{8}$, D. White$^{3}$, G. Woan$^{4}$}

\address{$^{1}$Cardiff University, Cardiff, CF24 3AA, United Kingdom }
\address{$^{2}$The University of Texas at Brownsville and Texas Southmost College, Brownsville, TX  78520, USA }
\address{$^{3}$The University of Sheffield, Sheffield S10 2TN, United Kingdom }
\address{$^{4}$University of Glasgow, Glasgow, G12 8QQ, United Kingdom }
\address{$^{5}$University of Maryland, College Park, MD 20742 USA }
\address{$^{6}$Departement Artemis,  Observatoire de la C\^ote d'Azur, CNRS, F-06304 Nice,  France }
\address{$^{7}$University of Wisconsin-Milwaukee, Milwaukee, WI  53201, USA }
\address{$^{8}$University of Birmingham, Birmingham, B15 2TT, United Kingdom }
\date{\today}

\begin{abstract}
The radio skies remain mostly unobserved when it comes to transient
phenomena.  The direct detection of gravitational waves will mark a major milestone of modern astronomy, as an entirely new window will open on the universe. Two apparently independent phenomena can be brought
together in a coincident effort that has the potential to boost both
searches. In this paper we will outline the scientific case that stands
behind these future joint observations and will describe the methods
that might be used to conduct the searches and analyze the data. The
targeted sources are binary systems of compact objects, known to be strong
candidate sources for gravitational waves. Detection of transients
coincident in these two channels would be a significant `smoking gun'
for first direct detection of gravitational waves, and would open up a
new field for characterization of astrophysical transients involving
massive compact objects.
\end{abstract}

\maketitle

\section{Introduction}
\label{sec:intro}

Many
potential sources of transient gravitational wave signals may emit
electromagnetic counterparts detectable by existing and planned astronomical instruments.
The coincident detection of an electromagnetic signal may provide
some of the most compelling evidence for the unambiguous direct detection of
gravitational waves, as well as provide important information on the
nature of the progenitor system.  Short, hard $\gamma$-ray bursts (GRB) provide a typical
example of such a scenario.  These are believed to be the electromagnetic signatures of
the coalescence of a compact binary system, consisting of two neutron stars (NS) or
a neutron star and a black hole (NS-BH).  GRBs have been used quite extensively
to trigger \gw searches for some time~\cite{Abbott:2007rh,Collaboration:2009kk}
and, indeed, a recent search for \gws from a GRB initially associated with the Andromeda galaxy
was able to confidently exclude a compact binary coalescence at the distance of M31 due to the absence
of significant \gw emission~\cite{GRB070201}, but did not exclude a binary coalescence event at larger distances.

Searching for \gws associated with a specific electromagnetic event provides several
advantages over a normal `all-sky' search.  Typically, an
all-sky search is performed over an entire science run, lasting weeks to
months. An electromagnetically {\it triggered search}, by contrast, is usually 
performed over a much shorter time window lasting a few to several hundred seconds and
there is often also very good directional information involved in the search. The smaller time window
allows for an increase in sensitivity due to a smaller number of instrumental
and terrestrial artifacts in the data, so one is able to tolerate signal detections
with lower significance than would be the case for a longer duration search.  In
addition, a sky location and gravitational wave form known {\it apriori}, coupled with a 
network of detectors, significantly reduces the parameter space of the search and
many instrumental artifacts which would otherwise mimic a gravitational wave
signal will simply be inconsistent with the given sky location.

Observations in radio astronomy have already had a significant impact on
the search for gravitational waves.  Most significantly, the accurate
timing of pulsars in radio has enabled searches for gravitational wave
emission from known pulsars \cite{known_pulsars}.  These radio
observations permit a significant reduction of the gravitational wave
parameter space, resulting in a more sensitive search.
Recently, the gravitational wave emission from the Crab pulsar has been
bounded to be significantly below the spindown limit \cite{crab}.  In
addition, observations of pulsar glitches \cite{Flanagan:2006,Buchner:2008} have prompted
searches for gravitational waves emitted at the time of the glitch~\cite{Clark:2007}.

In this paper, we advocate the extension of the joint radio and
gravitational wave search effort to include transient signals in the
radio band.  Until now, there have been no completely systematic searches for transient radio
signals but there are tantalising hints of a significant population of transients
\cite{Lazio:2009xe} which a new generation of radio telescopes and
arrays are ideally positioned to observe.  Given the nascent state of
the field, there is great uncertainty regarding the nature of the progenitor of 
many radio transients. Several of the proposed sources of radio transients are
also expected to be strong and, in some cases, well-modelled sources of 
gravitational waves.  The potential for serendipitous discovery
of new gravitational wave and radio sources, as well as the existence of 
theoretically modeled mechanisms for radio emission associated with known classes of
astrophysical objects, provide strong motivation for proposing a joint
gravitational wave and radio observation effort. On top of this, the astrophysical 
information encoded in the radio and gravitational waveforms will likely be complementary.
Thus, as with many multi-wavelength or multi-messenger observations,
combining the data from these two different observing channels will
enhance the astrophysical understanding of the source.

The paper is structured as follows: in section \ref{sec:telescopes} we
outline the observational capabilities both in radio and gravitational
waves; section \ref{sec:sources} introduces a number of proposed theoretical
models predicting both radio and gravitational wave emission from a
number of sources; in section \ref{sec:search}
we present the search methods that could be utilized in a joint GW-radio
search and finally section \ref{sec:discussion} provides a summary and
future prospects.  

\section{Search tools: radio telescopes and gravitational wave interferometers}
\label{sec:telescopes}

\subsection{Radio telescopes and recent radio transients survey activity}
\label{ssec:radio_tele}

Radio telescopes fall into two categories --- dishes and aperture
synthesis arrays. We begin by enumerating in Table \ref{tab:radioinst}
some of the key specifications of radio telescopes proposed for use in
relation to coincident searches and follow-up in more detail on
a number of previewed telescopes to be used in the very first stages of
the search.

\begin{table}[h!]
\begin{center}
\begin{tabular}{|l|l|l|l|l|}\hline
Instrument & Band & Type & Field of View & Slew Time\\ \hline \hline
LOFAR & 40-240\,MHz & Array & $30^\circ$ beam(s) & Software \\ \hline
ETA & 29-47\,MHz & Array & Two $30^\circ$ beams & Software \\ \hline
NRAO Green Bank & 1.15-1.73 GHz & Dish &
0.027 sq. deg.&  $18^\circ$/minute\\ \hline
ARECIBO & 312\,MHz - 10.2\,GHz & Dish &
15~arcmin~@~312\,MHz & $<16$\,min \\
 & & & 0.5~arcmin~@~10.2\,GHz & \\

\hline
\end{tabular}
\end{center}
\label{tab:radioinst}
\end{table}

\paragraph{Low Frequency Array (LOFAR)} 
LOFAR is a UHF antenna array recently commissioned by a Dutch consortium
lead by the Netherlands Institute for Radio Astronomy (ASTRON) and the
University of Groningen. The instrument has made its first observational
trials at the end of August 2009 and according to the latest news from
\cite{Garrett:2009gp} the first international observational effort has
just been completed. The instrument and the capabilities afforded by the
design are discussed elsewhere \cite{lofar_case}. Briefly, the design
calls for the deployment of 41 ground stations centered in the
Netherlands and further stations extending throughout western Europe.
Each ground station comprises of an array of sensors, including between
48 and 96 each of ``low-band'' and ``high-band'' antennae\footnote{International stations further from the core group in
the Netherlands will add more antennae.} having usable bandwidths of
30-80\,MHz and 120-240\,MHz respectively and a maximum sensitivity of
10\,mJy.  One of the key science projects of LOFAR is to search for
radio transients.  Potential sources include X-ray binaries, GRBs, SNe
and AGN. 

\paragraph{Eight-meter Transient Array (ETA)} 
The Eight-meter-wavelength Transient Array \cite{Patterson:2008ie} has been
constructed and operated by researchers at Virginia Tech.  This
instrument is designed specifically to detect low-frequency radio
transients, covering the band 29--47 MHz with full-bandwidth sampling.  Its
flexible signal processing system supports a number of modes by phasing
its individual dipole antennas, but it will typically be operated with
two 30-degree-wide synthesized beams to do a broad continuous search.

\paragraph{Green Bank NRAO} 
The Green Bank Telescope (GBT) is the world's largest fully steerable
radio telescope. GBT is located at the National Radio Astronomy
Observatory's site in West Virginia, USA. GBT is a 100-meter telescope
on a wheel-and-track design that allows the telescope to view the entire
sky above 5 degrees elevation.  

\paragraph{Arecibo}
The Arecibo radio telescope in Puerto Rico, USA, is the world's largest
and most sensitive radio telescope (312\,MHz - 10.2\,GHz and 0.5\,mJy
sensitivity).  It is part of the National Astronomy and Ionosphere
Center (NAIC) operated by Cornell University.  The telescope itself
consists of a 305 meter diameter fixed primary reflector, with a
suspended platform containing secondary and tertiary reflectors along
with various receivers.  The telescope can be pointed within $20^\circ$
of zenith by moving the suspended platform, with a slew rate of
$24^\circ$/minute in azimuth and $2.4^\circ$/minute in zenith angle.
The secondary and tertiary reflectors correct for spherical aberration.

Systematic surveys of the transient radio skies are expected to be performed in the near future at a greater rate than in the past \cite{Lazio:2009xe}. The unexpected results from such past surveys include discoveries of completely new radio sources (e.g. Rotating Radio Transients,~\cite{BurkeSpolaor:2009rm}). 
As an example, in the summer of 2007 Green Bank Telescope took a survey of the
northern sky at 350\,MHz which covered 12,000 sq degrees. This survey~\cite{Hessels:2007ct} was
called the drift-scan survey because it was done while the azimuth track
was being refurbished. Data from this survey has thus far uncovered 25
new pulsars including 5 new millisecond pulsars.  This data is still
being searched for new pulsars and radio transients. This shows the
shear diversity and abundance in new radio sources that can be uncovered
by doing a rather short but systematic survey and reveals the potential
of a multi-messenger search.

\subsection{Gravitational wave interferometers}
\label{ssec:gw_det}

A global network of gravitational wave
interferometers has now been constructed and is taking data.  The instruments
constituting this network
include the Laser Interferometry Gravitational Observatory (LIGO), which operates two
observatories in the USA~\cite{Abbott:2007kv}; the French-Italian Virgo detector~\cite{Virgo_status}, based in Italy; the
British-German GEO600 detector in Germany~\cite{GEO_status} and the TAMA300 decector in Japan~\cite{TAMA_status}.  Data
from these detectors has been acquired and analyzed over the past
decade.  These detectors have achieved or come close to their design
sensitivities, and an extended science run of the
LIGO, GEO and Virgo detectors was completed over 2005 to 2007 (known as S5 in
LIGO/GEO and VSR1 in Virgo).  The detectors achieved a strain
sensitivity of better than $10^{-22}/\sqrt{\mathrm{Hz}}$ at their most
sensitive frequencies (around $100$ Hz).  This can be translated into
sensitivities to various sources, for example the LIGO detectors in S5
were sensitive to optimally oriented and located (i.e. overhead the
detector) binary neutron star signals to a distance of $35$ Mpc, and hundreds of Mpc for
more massive compact binary coalescences.  For short-duration, narrow-band transients,
such as the \gw signal one may expect from core-collapse supernovae,
this sensitivity corresponds to a gravitational wave energy as low as
$10^{-8} M_{\odot}$ for galactic events and $0.1 M_{\odot}$ for event in
the Virgo cluster at 16 Mpc.

Following the S5/VSR1 run, the LIGO and Virgo detectors have been
technically upgraded to enhanced configurations, and the latest science run
(S6/VSR2) began in the summer of 2009, aiming at collecting data at better sensitivities than S5/VSR1.  Following this data taking
period, both LIGO and Virgo detectors will be upgraded to advanced
configurations with approximately ten times the strain sensitivity of
the initial detectors.  For sources distributed uniformly in volume,
this corresponds to a sensitivity to a thousand times as many sources.
In terms of energy, the sensitivity will be $\sim 10^{-10} M_{\odot}$
for galactic events.  The advanced detectors are expected to begin
aquiring scientific data by 2015. 
 
\section{Radio and Gravitational Wave Sources} 
\label{sec:sources}

Joint observation of gravitational waves and their radio afterglow
requires a mechanism for the prompt generation of a radio counterpart to
the gravitational wave signal. Furthermore, to avoid self-absorption by
the source, models yielding coherent radio emission are favoured.  The
prospects for detecting gravitational waves from a given progenitor
depend on the details of the underlying engine, which in many cases are
still uncertain.  To pursue a joint radio and GW analysis, one requires
a reliable estimate of the delay between the gravitational and radio
waves, given by the dispersion measure of the media in which the wave
travels.  There are several possible progenitors for emission in both
gravitational and radio waves, two of which are discussed below:
coalescing neutron star binaries and short hard GRB afterglows.  We
conclude this section with a brief disucssion of the effects of
dispersion on the radio signal. 

\subsection{Neutron Star Binaries}

Binary neutron stars are one of the most promising candidate for
gravitational wave sources.  Indeed, the observations of several binary
pulsars provide strong evidence for the emission of gravitational waves
from these systems \cite{weisberg:2004}, as well as an estimate of the
rate of such coalescences in the nearby universe \cite{Kalogera:2004tn, Kalogera:2004nt}.  The
waveform emitted by a coalescing neutron star binary system has been
calculated to great precision in the post-Newtonian formalism
\cite{Blanchet:2002av}.  Initial and enchanced detectors are sensitive
to the signal to tens of Mpc while the advanced detectors are sensitive
to hundreds of Mpc.  Several gravitational wave searches for coalescing
neutron star binaries have already been performed \cite{Collaboration:2009tt,
Abbott:2009qj}. At the time of writing, there has been no confirmed direct
detection of \gws using purpose-built detectors.  However, the 
upper limits obtained on the rate of binary coalescences are now approaching 
those predicted by astrophysical arguments.  The expected rate of such coalescences
observed in the advanced LIGO and Virgo network is expected to be tens
per year.

There are a number of models for the emission of radio
waves during the late stages of a compact binary inspiral phase or during their coalescence,
making these an ideal source for joint radio-GW searches.  We
discuss two classes of radio emission models below, based on the predicted emission mechanism.

\paragraph{Radio emission due to strong magnetic fields}

The first class of models
require one of the neutron stars to possess a large magnetic field
($10^{12} - 10^{15}$ G).  The orbital motion of the binary generates
time dependent magnetic fields and consequently induced electric and
magnetic fields.  The motion of these fields may then result in 
radio emission.

The model described in \cite{Lipunov:1996wf} assumes the binary
neutron star system is composed of stars with (approximately) equal
masses and radii in the final stages of inspiral.  One of the two NS
is required to be a magnetar, with magnetic field
$B\sim10^{12}-10^{15} \mathrm{G}$, with the second star's magnetic
field significantly weaker.  Their spins are neglected.  By
modelling the stars as perfect conducting spheres, it can be shown
that as the companion orbits in the magnetic field of the magnetar,
a magnetic dipole is induced in the companion and
dipolar radiation is emitted through Poynting
losses.  The expected maximum in source luminosity is of the order
of $L_{max}\sim 10^{41}$ erg/s. It is thought that, in
analogy with the pulsar model, a fraction of this energy will be
radiated in radio band.

In a second model \cite{Hansen:2000am}, the magnetar's companion is
assumed to be a   rapidly spinning recycled pulsar (with a period $P
\sim 1-100$ ms), with a low magnetic field.  The magnetar is a
non-recycled slow-spinning pulsar ($P \sim 10-1000$ s). As before,
the orbital and rotational motion of the companion result in an
induced dipolar electric field on its surface.  The majority of the
energy lost by the neutron star is converted into plasma, and later
radiated.  Given the lack of a complete theory for the emission, the
authors assume that $\epsilon \sim 0.1$ of the initial beam energy
is radiated in radio band at a reference frequency of 400 MHz (this
frequency and efficiency are chosen in analogy with radio pulsar
observations).  The observable flux would be of the order
$\mathrm{F_{\nu} \sim 2\,mJy}$ for a source placed at 100 Mpc.

\paragraph{Plasma excitation through relativistic magnetohydrodynamics}

It is well known (see
e.g.~\cite{Duez:2005sg,Duez:2005sf,Moortgat:2005fs,Moortgat:2004xz})
that, within relativistic magnetohydronamics (MHD), gravitational
waves will generically cause excitations of waves in the fluid.
Specifically, it will excite three wave modes in the fluid: Alfven
waves, fast and slow magnetosonic waves.  Thus, in
astrophysical situations with strong gravitational waves travelling
through strongly magnetized plasmas ~\cite{Moortgat:2004xz}, energy
can be transferred from the gravitational field to the plasma.
However, the MHD modes are initially excited at the same frequency
as the emitted gravitational waves.  The challenge then is to
determine whether there will be sufficient up-conversion to higher
frequencies that the energy might escape as electromagnetic
radiation.

In \cite{Moortgat:2004xz,Moortgat:2005fs} the authors argue that
this process could lead to an observable radio signal associated to
binary neutron star coalescences.  The inverse Compton scattering of
the MHD wave by a relativistic ouflow of secondary particles will
lead to the emission of radiation.  When the binary is close to face
on to the observer, this radiation will be observed at radio
frequencies, within the sensitive band of the future radio array detectors.
The nature of the signal predicted in ~\cite{Moortgat:2005fs} is an
incoherent burst of radio waves around 30 MHz and with a bandwidth
of 30 MHz with a duration of roughly 3 minutes for a source located
at 1\,Gpc. The predicted fluxes lie in the MJy region, but due to
the very efficient damping mechanisms predicted in parallel to this
model, the detected flux will most probably be much smaller, but
still within the sensitivity of LOFAR.

It is worth mentioning that these two phenomenological model cathegories do not exclude each other: radio emission due to the presence of a highly magnetized neutron star and its subsequently induced magnetic and electric fields is predicted to occur before the binary merger, wheres interactions of gravitational waves with the surrounding post-merger plasma and consequent MHD phenomenae will trigger a radio signal after the merger.

\subsection{Gamma ray bursts}

The coalescence of a binary neutron star or neutron star-black hole system
is still the main candidate progenitor of short hard gamma ray bursts (SHB).  Also, such mergers produce much stronger gravitational wave
emission than other predicted sources, making them some of the most promising candidates for gravitational-wave
detections with the first or second generation of ground-based detectors. The afterglows of SHB have been observed in different wavelenghts.   
Short hard GRBs are known for their weak afterglows, making it difficult to secure confirmation of the
progenitor. Results have been published on radio afterglows for short hard bursts but
the data shows only weak signals hours or days after the burst ~\cite{Ofek:2006pr,Soderberg:2006bn}.  

Several authors \cite{Moortgat:2003jh,Usov:2000yr,nakar07} have argued
in favour of a radio component of afterglows from short GRBs, namely a radio burst several minutes after the
observed GRB.  This radio burst is predicted to be a result of synchroton emission of the electrons in the post-merger plasma and is thought to have a flux on the order of mJy, which is within the sensitivity of current radio telescopes.  In
fact, a proposed discriminant \cite{nakar07} of baryon-dominated (as
opposed to magnetic-field-dominated) outflows is the presence of a radio
flare, stronger than the early optical afterglow, within the first half
hour after the burst.

Core collapse within massive stars is one of the most widely predicted sources of
transient gravitational and electromagnetic radiation.  This is the underlying
mechanism of supernovae, which occur a few times per century in galaxies
like our own.  At higher masses this collapse can produce long gamma-ray
bursts, which are observed at a rate of $10^{-7}\,\mathrm{yr}^{-1}$ per
galaxy, though the intrinsic rate is likely one or two orders of
magnitude higher due to beaming~\cite{Sadowski:2007dz}. However, the strength of
gravitational-wave emissions from supernovae is quite uncertain.
Optimistically it could be as high as $10^{-4}M_\odot c^2$ of energy
released as gravitational waves between 500 and 1,000 Hz
\cite{ott:201102}. Gamma-ray bursts may produce highly-beamed radio bursts within minutes of the gamma-ray burst~\cite{Usov:2000yr}, and supernovae in
general may produce electromagnetic afterglows starting hours after the
initial energy release.

\subsection{Dispersion in the intergalactic and interstellar media and Compton scattering}
Radio waves are strongly coupled to charged
particles and, therefore, are potentially subject to the effects of 
self-absorption in ionized material surrounding the source and to dispersion 
in the interstellar and intergalactic media (ISM and IGM, respectively). 
Self-absorption effects are more pronounced when the radio emission 
is incoherent, as with some of the above emission scenarios associated with
binary neutron star systems.

Following~\cite{pulsar_book}, a radio pulse traveling
in the ionised ISM is delayed over its propagation time through 
free space by a time $\Delta t_{\rm delay}$,
\begin{equation}
\Delta t_{\rm delay} = 4.1\,{\rm ms}\,{\rm DM}~\nu^{-2}_{\rm GHz},
\end{equation}
where $\nu$ is the observation frequency and DM is known as the dispersion measure.
This is the integral along the 
line-of-sight of the electron density between the the observer and the source:
\begin{equation}
{\rm DM} \equiv \int {\rm d}r~n_e(r),
\end{equation}
where $r$ is the distance to the source and $n_e(r)$ is the electron number
density at $r$.

Now, the IGM has a much lower electron number density than the ISM, but the
radio signal must travel a far greater distance through the IGM ($\sim 100$\,Mpc
for advanced LIGO) than through the ISM where the emission must propagate across
only $\sim 10$\,kpc for galaxies similar to the Milky Way. In the plane of
the Milky Way, the number density of electrons is, on average, about
$\mathrm{n_e=0.03\,cm^{-3}}$~\cite{thompson}.  So the dispersion
measure for 10\,kpc of ISM equivalent to the disc of
the milky way is $\mathrm{\sim 300\,cm^{-3}\,pc}$.  The expected dispersion
measure contribution for intergalactic distances is $\mathrm{DM} \approx 
100$\,pc\,cm$^{-3}$~\cite{skadoc,palmer} and so the dispersion due to the intergalactic and
interstellar media along the signal path are comparable.  The time
delay due to dispersion for a 1\,GHz radio signal is estimated to be
less than 4\,s for sources within range of advanced LIGO \cite{laz}.
Taking this number and adding a component for dispersion in the
interstellar medium, we can estimate that dispersion delays of order
a few seconds for sources embedded in Milky Way-like galaxies at
distances of order a few 100\,Mpc may be expected.
Since the time delay is inversely proportional to the second power of frequency,
lower-frequency signals may be delayed by many minutes; however, the
time of emission may still be inferred from a broadband signal by
extrapolating the delay-vs.-frequency function to infinite frequency.
We therefore retain the benefits of a triggered search.

In the case of short hard GRB radio afterglows, apart from dispersion, the radio waves emited by such bright sources may suffer from induced Compton scattering within the source, phenomenon that will cause a significant dampening of the signal. Detailed in~\cite{Macquart:2007kv}, the induced Compton scattering is the main limiting factor when the region around the progenitor is not dense but when one still considers the scattering effect of a tenuous circumburst interstellar medium. The presence or absence of a radio emission provides an excellent constraint on the Lorentz factor of the GRB outflow during the very early stages of its outburst, hence providing information on the energetics of the progenitor and its nature.

\section{Joint radio-GW searches}
\label{sec:search}

There are two ways in which the coincident detection of a radio--GW
event can be made: either by following up radio transients in existing
gravitational wave data, starting with existing radio transients
detected during the past and present science runs, or by using the
prompt detection and localization of gravitational waves as initial
trigger and {\it alerting} the radio telescopes to point in the
direction where the gravitational wave was observed.  We will discuss
each of these in turn.

\subsection{Follow-up of radio transients in archived gravitational wave data}

As we have argued earlier, performing an electromagnetically triggered
search of gravitational wave data has several advantages over the all
sky, all time searches.  The external trigger allows for a significant
reduction in the data to be searched, both by restricting the time
duration and also the sky position.  This reduction in parameter space leads
to a corresponding increase in the sensitivity of the search.  Given the
theoretical models presented in the previous section, there is a clear
motivation for performing a follow-up observation in gravitational wave
data of radio triggers.  Gravitational wave data is routinely archived and also, 
there is no inherent time restriction in performing the search. Indeed, if there
are radio transients identified at times that overlap previous
gravitational wave detector science runs, it is possible and much desired to
search the gravitational wave data around these times.

An outstanding challenge is to obtain a better understanding of the
relative timing of the radio and GW signals.  Once the GW time window is
greater than a few hours, much of the benefit of performing a follow-up style
search is lost.  Thus, it is imperative that we improve our
understanding of the various models presented above to obtain good
estimates of timing differentials between GW and radio signals.  An
interesting aspect of the follow-up of radio triggers is that for each event
we will have an estimate of the dispersion measure.  By measuring the
dispersion, it should be possible to correct for any time delay of the
radio signal.  Furthermore, this should provide an independent
measure of the distance, which could be compared with any GW
observations.

The follow-up searches begin with a list of radio transients; for each
one a GPS time, the duration, the energy of the burst, the dispersion
measure and sky location are recorded. For each event, we advocate the use all available
LIGO/Virgo data at the time of the event to follow-up these events.
Given the source models discussed in section \ref{sec:sources} we
propose the following gravitational wave searches:

\begin{itemize}

\item Search for compact binary coalescence.  There is an argument to
focus this search on binary neutron star signals.  It should be straightforward to apply
a very similar search method to that used for searching for
gravitational waves associated to short GRBs \cite{Abbott:2007rh}.  Although
there are fewer models predicting radio emission from neutron star-black
hole and black hole-black hole binaries, it is straightforward to extend
the search to include these systems as well.  Interestingly, in the absence of a
detection, it should be possible to set a lower limit on the distance to
the source assuming that it is a binary merger. It should then be
possible to compare this limit to the distance inferred from the
dispersion measure; we may be able to say with some confidence that a given
radio burst was {\em not} caused by a binary merger.

\item Search for unmodeled bursts of gravitational waves.  When the radio
burst is well localized in the sky, it should be straightforward to make
use of the same methodology as has been previously applied in the search
for gravitational waves associated with GRBs \cite{Collaboration:2009kk}.  Some
radio antennae, for example the ETA radio array, will mainly operate in
a wide-area burst-search mode and therefore will not provide a good sky
localization.  In this case, the simplest search would be a coarse time
and sky location coincidence between radio and gravitational wave
triggers from a standard excess-power style, all sky burst search.

\end{itemize}

\subsection{Gravitational wave events followed-up by radio observations}

Gravitational wave antennae have broad-lobed antenna patterns covering
tens of degrees on the sky per instrument. Searching for transients that
excite a network of at least three spatially separated interferometers,
reconstructing the source position, and finally using astronomical
instruments to image the region of the reconstructed gravitational wave
source utilizes the best features of gravitational wave detectors. That
is, they look at a broad patch of the sky but, by utilizing a network of
detectors, will still allow us to estimate the source position to much higher
precision than the field of view.  However, the intrinsic pointing
accuracy of the LIGO/Virgo gravitational wave network is still on the
order of tens of square degrees \cite{Fairhurst:2009tc}.  This
localization accuracy is similar for signals close to threshold in both
the initial and advanced detector networks, but could be significantly
improved for the loudest signals in advanced detectors.  

For following up gravitational wave events, aperture synthesis arrays
such as LOFAR and ETA offer some key advantages.  Signals from multiple
antennae are correlated to synthesize a beam far narrower than the
antenna pattern of a single antenna, which may be as simple as a dipole.
The parameters of the correlation may be tuned to allow beams as wide as
30 degrees with resolving power as good as 0.5 arc seconds. Reaction
time of the instrument is dependent on the software driving the
correlator. A key difference between aperture synthesis arrays of radio
telescopes and a gravitational wave detector network is that the
sampling rate in the radio precludes archiving data for more than a few
seconds, so that there is no look-back capability.  Thus, the key
challenge for this search is the rapid analysis of the gravitational
wave data, to allow for timely pointing of radio arrays.

\section{Summary}
\label{sec:discussion}

In this paper we have presented the case for joint gravitational wave
and radio analysis between data from the gravitational wave detectors
available to the LSC/Virgo collaborations and a variety of radio
astronomical instruments.  It encompasses a variety of projects, some of
which are minor extensions of existing projects in the optical or gamma
ray observations, whilst most of the others make use of existing
software or collaborations. 

A series of scientific factors motivate such a coincident search: it is
a logical evolution of the previously started gamma ray coincident searches, extended to the radio band; 
there are a series of
published theoretical models that predict radio events preceded or
followed by gravitational wave emissions; computationally speaking,
the data analysis pipelines are already in place for such a task.
Although, in broad lines, the proposed observational and data analysis
mechanisms are similar to the ones that already exist as multi-messenger
projects and are a logical evolution of the multi-messenger concept
itself in exploring all the available electromagnetic bands, a
coincident detection of radio and gravitational wave transients poses a
set of very particular issues to this type of search. One of these
issues is that the theoretical models considered are far from being
confirmed by astronomical observations; the diversity of the models
suggests a very broad approach from the side of the theorists and a
burden for us, the astronomical observers, in confirming or
disconfirming the models in an environment where choosing a particular
model may increase the sensitivity of the whole search. Another issue is
that, even if most of the computational tools are in place, the initial search
parameters are still to be decided, e.g. the time window around a radio
transient that should be searched over in gravitational wave data, the
dispersion measure based on the IGM and ISM number density, the mass
distribution and populations of the compact objects thought to be the
sources of both radio and GW transients.  Thus, there is much to be explored in the near future.

\section*{Acknowledgements}

This research was made possible through support form the Science and
Technology Facilities Council and the Royal Society in the United Kingdom
as well as the National Science Foundation in the United States of
America. The authors would also like to thank their colleagues within the LIGO and Virgo collaborations for discussions and subsequent suggestions and comments.

\section*{References} 

\bibliographystyle{unsrt}
\bibliography{radio}

\end{document}